\documentclass[12pt,english]{article}
\usepackage[T1]{fontenc}
\usepackage{a4wide}
\usepackage{babel}
\usepackage{psfrag}
\usepackage{graphicx}

\newcommand{\be}{\begin{equation}}
\newcommand{\ee}{\end{equation}}
\newcommand{\bea}{\begin{eqnarray}}
\newcommand{\eea}{\end{eqnarray}}

\newcommand{\ba}{\begin{array}}
\newcommand{\ea}{\end{array}}
\newcommand{\vs}[1]{\vspace{#1 mm}}

\def\bbox{{\,\lower0.9pt\vbox{\hrule \hbox{\vrule height 0.2 cm
\hskip 0.2 cm \vrule height 0.2 cm}\hrule}\,}}
\newcommand{\dsl}{\pa \kern-0.5em /}

\newcommand{\pa}{\partial}

\font\mybb=msbm10 at 12pt
\def\bb#1{\hbox{\mybb#1}}

\begin{document}
\input{epsf}

\topmargin 0pt
\oddsidemargin 0mm

\renewcommand{\thefootnote}{\fnsymbol{footnote}}
\begin{titlepage}

\begin{flushright}
KCL-TH-00-51\\
September 2000

\end{flushright}
\vs{15}

\begin{center}
{\Large\bf Novel Domain Wall and Minkowski Vacua of D=9 Maximal $SO(2)$ Gauged Supergravity}
\vs{10}

{\large P.M. Cowdall \footnote{
e-mail: {\tt pcowdall@mth.kcl.ac.uk }}}

\vs{5}
{\it Department of Mathematics, King's College\\ The Strand, 
London, UK, WC2R 2LS}

\end{center}
\vs{10}
\centerline{{\bf Abstract}}
We show that a generalised reduction of D=10 IIB supergravity leads, 
in a certain limit, to a maximally extended SO(2) gauged supergravity in D=9.
We show the scalar potential of this model allows both Minkowski and a new
type of domain wall solution to the Bogomol'nyi equations. We relate these
vacua to type IIB D-branes.

\end{titlepage}
\newpage 
\renewcommand{\thefootnote}{\arabic{footnote}}
\setcounter{footnote}{0}

\newpage
\section{Introduction}

In recent years gauged supergravities have enjoyed a revival of interest. 
These models have come under renewed investigation because it is now clear 
they play an important role in the dynamics of superstring theories and 
M-theory through the conjectured AdS/CFT correspondence and
its generalisations \cite{malda,BST}.

The most studied of these holographic correspondences is that between type
IIB superstring theory in the background $AdS_5{\times}S^5$ and super 
Yang-Mills theory in four dimensions with gauge group $SU(N)$ for $N$ large.
When the string coupling $g_s$ becomes small but $g_sN$ remains large, the 
radius of curvature of $AdS_5{\times}S^5$ becomes much larger than the string
length and hence the supergravity approximation is justified. The relevant 
supergravity would be type IIB supergravity compactified on $S^5$, whose
massless modes constitute D=5 maximal $SO(6)$ gauged supergravity
\cite{GRW}.

Gauged supergravities also arise in brane-world scenarios \cite{RS} in which the 4-D
fields of the Standard Model are localised on a 3-brane domain wall living
in five dimensions. The bulk geometry is asymptotically $AdS_5$ and although
the fifth dimension is non-compact, the zero mode of the graviton is trapped
by the wall and four dimensional Newtonian gravity is correctly reproduced on the wall. As $AdS_5$ is a solution of 5-D maximal gauged supergravity, which is believed to be a consistent truncation of IIB supergravity
\cite{CLPST,Gunmarc,KRvN}, attempts have been made to
embed the braneworld models in M/Superstring theory \cite{LOSW,ELPP,HW,EdWitten}.

The purpose of this paper is to study maximally extended, gauged supergravity
in nine dimensions. This model is interesting from the point of view of
D=9 being the highest dimension in which the construction of a gauged
supergravity is possible. For D=11 and D=10 IIA supergravities the
$R$-symmetry groups of the superalgebras are trivial. In the D=10 IIB
case although the $R$-symmetry group is $SO(2)$ there are no vector
fields in the multiplet available to act as gauge fields. 

A natural place to begin then is with the D=9 maximal ($N$=2) ungauged model 
\cite{KK,BHO}. 
The bosonic field content is a graviton, a 3-form potential, a pair of
2-forms, 3 vectors and 3 scalars. The model has a global $SO(2)$ invariance
under which the 2-forms transform as a doublet as do 2 of the vectors. 2
of the scalars parametrise the coset $Sl(2,\bb{R})/SO(2)$. The 3-form
is a singlet as is one of the scalars and the remaining vector, in a particular choice of field 
variables \cite{KK}. The $R$-symmetry group of the $N$=2 superalgebra is $SO(2)$ so it
seems clear that upon gauging the single vector must become the gauge field.
However, because
there are doublets of vectors and 2-index antisymmetric tensor potentials
also transforming under the $SO(2)$, one encounters similar problems that have arisen previously in the gauging of,
for example, maximal D=6 supergravity \cite{cownew}. The problem is that these doublets 
of vectors and 2-index antisymmetric tensor potentials have their own 
antisymmetric tensor gauge invariance which is destroyed upon replacement of 
ordinary derivatives by $SO(2)$ covariant derivatives in their kinetic terms.
Hence these doublets of fields would propagate an incorrect number of degrees
of freedom compared to that required by supersymmetry. So how does one couple
these vectors and 2-forms to the $SO(2)$ gauge field? The correct mechanism 
is to allow the doublet of 2-forms to eat the doublet of 1-forms and thus 
become massive. The antisymmetric tensor gauge invariance no longer exists 
but is not required as the doublet of massive 2-forms propagates the correct 
number of degrees of freedom. Now however the ordinary derivative can simply 
be replaced by a covariant derivative in the kinetic term for the massive 
2-forms in order to effect the gauging.

One possible method of constructing a D=9 gauged supergravity is via a
compactification on $S^2$ of D=11 supergravity. To see this consider the D=11
solution of three M5-branes intersecting on a string. This is a configuration
preserving 1/8 of the supersymmetry. The
`near horizon geometry' of such a configuration has been shown to be
$AdS_3{\times}{\bb{E}}^{6}{\times}S^2$ where the supersymmetry is enhanced by a
factor of 2 \cite{boonstra}. By `near horizon geometry' it is meant the geometry of the solution in the asymptotic region near the core of the branes where spacelike geodesics can be continued indefinitely. The existence of this solution implies that a reduction of D=11
supergravity on $S^2$ is possible, and, assuming the truncation to massless
Kaluza-Klein modes is consistent, one would expect to obtain a gauged
supergravity whose gauge group is a subgroup of the isometry group of $S^2$
i.e. $SO(3)$ \cite{cowpaper}
 
The method we will employ to obtain the bosonic sector of maximally extended
$SO(2)$ gauged
supergravity in D=9, is a generalised dimensional reduction of D=10 type IIB
supergravity using the global $Sl(2,\bb{R})$ symmetry. This has been done
previously and one obtains a massive supergravity with 3 mass parameters,
a global $Sl(2,\bb{R})$ and a local $SO(2)$ \cite{LLP,MO}. Here we review this reduction,
pointing out some features specific to the generalised reduction. In
particular, how the normally trivial local $U(1)$ symmetry, related to
general coordinate transformations in the compact dimension (for which
the Kaluza-Klein vector is the gauge field) becomes non-trivial after a
generalised reduction. The novelty of the present discussion is the observation that there exists a limit of this massive,
gauged supergravity in which two of the mass parameters can be set to zero
\footnote{In fact all three mass parameters can be put to zero giving the
maximally supersymmetric ungauged D=9 supergravity \cite{KK}.} giving the $SO(2)$
gauged supergravity of interest. We examine the scalar potential of this model
and find, rather surprisingly, that there is a single critical point leading
to a nine dimensional Minkowski ground state with no linear dilaton. We also
confirm the observation of \cite{LLP} that there are no supersymmetric single scalar
domain walls but the Bogomol'nyi equations do however allow domain walls with two scalars active. The functional dependence of these scalars on the
transverse distance from the wall is of a type not previously encountered.

The paper is set out as follows. In section 2 we review the generalised
reduction of type IIB supergravity to D=9 familiarising the reader with the notations used in later sections. In section 3 we take a limit to
obtain the bosonic sector only of the $SO(2)$ gauged supergravity. In section 4 we analyse the possible vacua of this model. In section 5 we briefly discuss the further
reduction of D=9 $SO(2)$ supergravity to lower dimensions and finally in
section 6 we present our conclusions.

\section{Generalised Reduction of IIB Supergravity}

Here we perform the generalised reduction of IIB supergravity to D=9. The
field equations of D=10 IIB supergravity cannot be derived from a covariant
action \cite{IIBeqns}. This is due to the self-dual 5-form field strength. However, as
shown in \cite{BBO}, one can write down a covariant action whose field equations
become those of IIB supergravity when the self-duality constraint is taken
into account. The form of the bosonic sector of this action with a manifest
$Sl(2,\bb{R})$ global symmetry is \cite{LLP,MO}:
\be \label{IIBaction}
I_{10} = \int d^{10}x\,e\bigg\{ R+{1\over4}tr(\partial{\cal{M}}\partial
{\cal{M}}^{-1})-{1\over12}{\bf{\cal{H}}}_{3}^{T}{\cal{M}}{\bf{\cal{H}}}_{3}\bigg\} \\
-{1\over4}\, \int
H_{5}\wedge\star{H}_{5}+
H_{5}\wedge{\bf B}_{2}^{T}\eta{\bf H}_{3} 
\ee
where for convenience we have used form notation in the terms
involving
$H_5$, and where    
\be
{\bf{\cal{H}}}_{3}\equiv{\bf H}_{\mu\nu\rho}={\pmatrix {H_{\mu\nu\rho}^{(1)} \cr H_{\mu\nu\rho}^{(2)}}}=
3{\pmatrix {\partial_{\mu} B_{\nu\rho}^{(1)} \cr \partial_{\mu} B_{\nu\rho}^{(2)}}}=3\partial_{\mu}
{\bf B}_{\nu\rho}, \qquad H_5 = dB_{4}+{1\over2}{\bf B}_{2}^{T}\eta{\bf H}_{3}.
\ee
The two scalars $\chi$ and $\phi$ parametrise an $Sl(2,\bb{R})/SO(2)$
coset. The symmetric $Sl(2,\bb{R})$ matrix $\cal{M}$ is defined by
\be
{\cal{M}}=e^{\phi}{\pmatrix {|{\tau}|^{2} & {\chi} \cr {\chi} & 1}}
\ee
and satisfies ${\cal{M}}^{-1}=\eta{\cal{M}}{\eta}^{T}$. $\tau=\chi+
ie^{-\phi}$ is a complex scalar field and the matrix $\eta$ is the generator
of $SO(2)$ given by
\be
\eta = {\pmatrix{{}0 & 1 \cr -1 & 0}} = -{\eta}^{-1}=-{\eta}^{T} 
\ee
which is preserved by all $Sl(2,\bb{R})$ matrices {\it i.e.}
$\ {\Lambda}^{-1}=\eta{\Lambda}^{T}{\eta}^{T}\quad\forall \; \Lambda \in 
Sl(2,\bb{R})$.

In a generalised dimensional reduction certain fields {\it are} allowed to
depend on the compactification coordinate. The dependence is dictated by how
the fields transform under the global symmetry being used, and is such that
the resulting lower dimensional lagrangian is independent of the
compactification coordinate. The $Sl(2,\bb{R})$ transformations of the fields
are
\be
{\cal{M}}\rightarrow\Lambda{\cal{M}}{\Lambda}^{T}, \qquad {\bf B}_{2}
\rightarrow ({\Lambda}^{T})^{-1}{\bf B}_{2}
\ee
where $\Lambda \in Sl(2,\bb{R})$. Hence the ans\"atze we choose for the
reduction are \footnote{Hatted quantities are ten dimensional, unhatted
are nine dimensional. $z$ is the compactification coordinate and $x$ are
the D=9 coordinates.}
\bea \label{genredansatz}
{\hat {\cal{M}}}(x,z) & = &  {\lambda}^{T}(z){\cal{M}}(x){\lambda}(z) 
\label{Mansatz}
\\
{\hat{\bf B}}_{2}(x,z) & = & {\lambda}^{-1}(z)[{\bf B}_{2}(x)-{\bf B}_{1}
(x){\cal{A}}_{1} + {\bf B}_{1}(x)(dz+{\cal{A}}_{1})] \label{Bansatz}.
\eea
$\lambda(z) = e^{zC}$ and the matrix $C$ is an element of the Lie
algebra of $Sl(2,\bb{R})$ given by
\be
C = {1\over2}\sum_{i=1}^{3}m_{i}\,T_{i} = {1\over2}
{\pmatrix{m_{1} & m_{2}+m_{3} \cr  m_{2}-m_{3} & -m_{1}}}
\ee
where
\be
T_{1}={\sigma}^{3}=\pmatrix{1 & {}0 \cr 0 & -1}, \qquad
T_{2}={\sigma}^{1}=\pmatrix{0 & 1 \cr 1 & 0}, \qquad
T_{3}=i{\sigma}^{2}=\pmatrix{{}0 & 1 \cr -1 & 0}=\eta
\ee
are the generators of $Sl(2,\bb{R})$ and the $m_i$'s are constants with the dimensions of mass. For the metric we use the standard
ans\"atz, namely
\be \label{metricansatz}
dS_{10}^{2} = e^{2\alpha\varphi}dS_{9}^{2} + e^{2\beta\varphi}
(dz+{\cal{A}}_{1})
\ee
or in terms of the vielbein
\be
{\hat e}_{\hat{\mu}}{}^{\hat{a}} = \pmatrix{ e^{\alpha\varphi}{e}_{\mu}
{}^{a} & e^{\beta\varphi}{\cal{A}}_{\mu} \cr 0 & e^{\beta\varphi}}.
\ee
$\beta=-7\alpha$ and $\alpha={1\over4{\sqrt7}}$ are chosen so that
the Einstein-Hilbert term thus reduces in the canonical way
\be
{\hat e}{\hat R}=e\bigg\{ R-{1\over2}(\partial\varphi)^{2}-{1\over4}
e^{-{4\over{\sqrt7}}\varphi}{\cal{F}}_{2}^{2}\bigg\}
\ee
where ${\cal{F}}_{2}=d{\cal{A}}_{1}$ is the field strength of the
Kaluza-Klein vector. 
\subsection{Reduction of Scalars}

We now reduce the scalar kinetic term which takes the form
\be
\label{scalars1}
{1\over4}{\hat e}\;tr\bigg[({\partial}_{{\hat{\mu}}}{\hat{\cal{M}}})
({\partial}_{{\hat{\nu}}}{\hat{\cal{M}}}^{-1}){\hat g}^
{{\hat{\mu}}{\hat{\nu}}}\bigg].
\ee
The inverse metric is given by
\be
{\hat g}^{{\hat{\mu}}{\hat{\nu}}}=\pmatrix{e^{-2\alpha\varphi}g^{\mu\nu}
& -e^{-2\alpha\varphi}{\cal{A}}^{\mu} \cr -e^{-2\alpha\varphi}{\cal{A}}^{\nu}
& e^{-2\alpha\varphi}{\cal{A}}_{\mu}{\cal{A}}^{\mu}+e^{-2\beta\varphi}}
\ee
which upon substitution in (\ref{scalars1}) gives
\bea \label{scalars2}
\!\!\!\!
{1\over4}e^{2\alpha\varphi}e\;tr\bigg\{&& \!\!\!\!\!\!\!\!\!
({\partial}_{\mu}{\hat{\cal{M}}})({\partial}_{\nu}{\hat{\cal{M}}}^{-1})
(g^{\mu\nu}e^{-2\alpha\varphi})-
({\partial}_{\mu}{\hat{\cal{M}}})({\partial}_{z}{\hat{\cal{M}}}^{-1})
({\cal{A}}^{\mu}e^{-2\alpha\varphi}) \nonumber\\ 
&& \!\!\!\!\!\!\!\!\!\!\!\!\!\!
-({\partial}_{z}{\hat{\cal{M}}})({\partial}_{\nu}{\hat{\cal{M}}}^{-1})
({\cal{A}}^{\nu}e^{-2\alpha\varphi})+
({\partial}_{z}{\hat{\cal{M}}})({\partial}_{z}{\hat{\cal{M}}}^{-1})
[e^{-2\alpha\varphi}{\cal{A}}_{\mu}{\cal{A}}^{\nu}+e^{-2\beta\varphi}]
\bigg\}.
\eea
In standard dimensional reductions, only the first term of this expression
would contribute to the lower dimensional lagrangian. In a generalised
dimensional reduction however we must be careful as 
${\partial}_{z}{\hat{\cal{M}}}
\neq0$. The first term in (\ref{scalars2}) trivially reduces as
\be
tr\bigg[({\partial}_{\mu}{\hat{\cal{M}}})({\partial}_{\nu}{\hat{\cal{M}}}
^{-1})g^{\mu\nu}\bigg]=tr\bigg[({\partial}{\cal{M}})({\partial}
{\cal{M}}^{-1})\bigg].
\ee
The terms in ${\partial}_{z}{\hat{\cal{M}}}$ reduce as
\bea
tr\bigg[({\partial}_{\mu}{\hat{\cal{M}}})({\partial}_{z}{\hat{\cal{M}}}^{-1})
{\cal{A}}^{\mu}\bigg] &=& 
-tr\bigg[({\partial}_{\mu}{\cal{M}})(C{\cal{M}}^{-1}+{\cal{M}}^{-1}C^{T})
{\cal{A}}^{\mu}\bigg] \nonumber\\
tr\bigg[({\partial}_{z}{\hat{\cal{M}}})({\partial}_{\nu}{\hat{\cal{M}}}^{-1})
{\cal{A}}^{\nu}\bigg] &=&
{\ }tr\bigg[(C^{T}{\cal{M}}+{\cal{M}}C){\cal{A}}^{\mu}({\partial}_
{\mu}{\cal{M}}^{-1})\bigg] \nonumber\\
tr\bigg[({\partial}_{z}{\hat{\cal{M}}})({\partial}_{z}{\hat{\cal{M}}}^{-1})
\bigg]
&=& 
-2\;tr\bigg[C^{2}+C^{T}{\cal{M}}C{\cal{M}}^{-1}\bigg]
\eea
where we have used (\ref{Mansatz}). We can now define covariant derivatives
as
\bea
D{\cal{M}} &=& {\partial}{\cal{M}}-(C^{T}{\cal{M}}+{\cal{M}}C){\cal{A}}_{1}
\nonumber\\
D{\cal{M}}^{-1} &=& {\partial}{\cal{M}}^{-1}+(C^{T}{\cal{M}}^{-1}
+{\cal{M}}^{-1}C^{T}){\cal{A}}_{1} 
\eea
and hence the scalar kinetic terms become the following after generalised
reduction to D=9:
\be
{1\over4}\,{\hat e}\,tr\bigg[{\partial}{\hat{\cal{M}}}
{\partial}{\hat{\cal{M}}}^{-1}\bigg] = 
{1\over4}\,e\,tr\bigg[D{\cal{M}}D{\cal{M}}^{-1}\bigg]- e{\cal{V({\cal{M}})}}
\ee
where 
\be
{\cal{V({\cal{M}})}}={1\over2}\,e^{{4\over{\sqrt7}}\varphi}\;tr\bigg[
C^{2}+C{\cal{M}}^{-1}C^{T}{\cal{M}}\bigg].
\ee
Thus we see that generalised reduction very naturally generates a scalar
potential and covariant derivatives. We now turn to the reduction of the
doublet of 2-forms.
\subsection{Reduction of 2-Form Potentials}

The ans\"atz for the reduction of the 2-forms was given in (\ref{Bansatz}).
Taking the exterior derivative generates an extra term not present in
ordinary reduction and there is also a common factor of $e^{-zC}$
appearing through the transformation law of ${\hat {\bf B}}_{2}\ ${\it i.e.}
\be
d{\hat {\bf B}}_{2}(x,z) = -Ce^{-zC}{\bf B}_{2}dz + e^{-zC}d{\bf B}_{1}
(dz+{\cal{A}}_{1}) + e^{-zC}(d{\bf B}_{2}-d{\bf B}_{1}{\cal{A}}_{1}).
\ee
In terms of field strengths we have
\be \label{H3ansatz}
{\hat {\bf H}}_{3}(x,z) = e^{-zC}{\bf H}_{3}(x) + e^{-zC}{\bf H}_{2}(x)
(dz+{\cal{A}}_{1})
\ee
where the D=9 field strengths are defined as
\be 
{\bf H}_{2}(x)=d{\bf B}_{1}-C{\bf B}_{2}, \qquad
{\bf H}_{3}(x)=d{\bf B}_{2}-{\bf H}_{2}\wedge{\cal{A}}_{1}.
\ee
Note how in the reduction ${\hat{\bf H}}_{3}$ splits into ${\bf H}_{3}$
 and ${\bf H}_{2}$ where ${\bf H}_{3}$ is given by the usual expression
involving a Chern-Simons modification. However, with the generalised
ans\"atz ${\bf H}_{2}$ also develops a modification. As we will see
later, this allows ${\bf B}_{2}$ to eat ${\bf B}_{1}$ and thus
${\bf H}_{3}$ becomes a covariant derivative. The kinetic term for
${\hat{\bf H}}_{3}(x,z)$ therefore reduces to
\be
-{1\over12}\,{\hat e}\,{\hat{\bf{\cal{H}}}}_{3}^{T}{\hat{\cal{M}}}
{\hat{\bf{\cal{H}}}}_{3} = 
-{1\over12}\,e\,e^{-{1\over\sqrt7}\varphi}{\bf{\cal{H}}}_{3}^{T}{\cal{M}}
{\bf{\cal{H}}}_{3}-{1\over4}
\,e\,e^{{3\over\sqrt{7}}\varphi}{\bf{\cal{H}}}_{2}^{T}{\cal{M}}
{\bf{\cal{H}}}_{2}. 
\ee
We now see how the $z$ dependence of the ans\"atz is such as to ensure
the lower dimensional lagrangian remains independent of $z$. Finally
it remains to deal with the terms involving the self-dual 5-form.
\subsection{Reduction of the 5-Form}

The ans\"atz for the 4-form potential is
\be \label{B4ansatz}
{\hat B}_{4}(x,z) = B_{4}(x)-B_{3}(x){\cal{A}}_{1}+B_{3}
(dz+{\cal{A}}_{1}).
\ee
The self-dual field strength is given by
\be
{\hat H}_{5}(x,z) = d{\hat B}_{4}+{1\over2}{\hat{\bf B}}_{2}^{T}\eta
{\hat{\bf H}}_{3}.
\ee
By using (\ref{B4ansatz}), (\ref{Bansatz}) and (\ref{H3ansatz}) it is
straightforward to show ${\hat H}_{5}(x,z)$ reduces as
\be \label{H5ansatz}
{\hat H}_{5}(x,z) = H_{5}(x) + H_{4}(x)(dz+{\cal{A}}_{1})
\ee
where $H_{4}(x)$ and $H_{5}(x)$ are defined by
\bea
H_{5} &=& dB_{4}-dB_{3}{\cal{A}}_{1}+{1\over2}({\bf B}_{2}^{T}+
{\cal{A}}_{1}{\bf B}_{1}^{T}){\eta}{\bf H}_{3} \nonumber\\
H_{4} &=& dB_{3}-{1\over2}{\bf B}_{1}^{T}{\eta}{\bf H}_{3}
+{1\over2}({\bf B}_{2}^{T}+
{\cal{A}}_{1}{\bf B}_{1}^{T}){\eta}{\bf H}_{2}
\eea
and use has been made of the property 
\be \label{sl2r}
e^{-z{C}^{T}}{\eta}\,e^{-zC}={\eta} \qquad\forall\;e^{-zC}\,\in\,
Sl(2,\bb{R}).
\ee
Hence the kinetic term for the 4-form potential dimensionally reduces to
\be
-{1\over4}\,{\int}_{\!\!\!\!{{\cal{M}}_{10}}}{\hat H}_{5}\wedge{\star}{\hat H}_{5}
=-{1\over4}\,
{\int}_{\!\!\!\!{{\cal{M}}_{9}}}e^{-{2\over\sqrt7}\varphi}
{H}_{5}\wedge{\star}{H}_{5}+e^{{2\over\sqrt7}\varphi}
{H}_{4}\wedge{\star}{H}_{4}.
\ee
The ten dimensional Chern-Simons term takes the form 
\be
I_{10}^{cs}=-{1\over4}\,{\int}
{\hat H}_{5}\wedge{\bf{\hat B}}_{2}^{T}\eta{\bf{\hat H}}_{3}.
\ee
The reduction of this term is performed using equations (\ref{H5ansatz}), (\ref{Bansatz}) and (\ref{H3ansatz}) and
the property (\ref{sl2r}). It is straightforward to show that the
resulting nine-dimensional Chern-Simons term is given by
\be
I_{9}^{cs}=-{1\over4}\,{\int}
H_{5}\wedge\bigg[({\bf B}_{2}^{T}+{\cal{A}}_{1}{\bf B}_{1}^{T}){\eta}
{\bf H}_{2}-{\bf B}_{1}^{T}{\eta}{\bf H}_{3}\bigg]
-H_{4}\wedge\bigg[({\bf B}_{2}^{T}+
{\cal{A}}_{1}{\bf B}_{1}^{T}){\eta}{\bf H}_{3}\bigg].
\ee

As mentioned at the beginning of section 2, the field equations of 
(\ref{IIBaction}) must be supplemented with the self-duality constraint
on ${\hat H}_{5}$ in order to be equivalent to the field equations
of type IIB supergravity. One handles the self-duality constraint in
nine dimensions by first dualising $B_{4}$ to ${\tilde B}_{3}$ by the
usual methods and then identifying ${\tilde B}_{3}$ with $B_{3}$.
The reader is invited to see \cite{MO} for a more complete discussion of this point. The result is
a contribution to the D=9 action involving only a 3-form potential
\bea \label{Lcs} 
I_{9}=
{\int}-{1\over2}e^{{2\over\sqrt7}\varphi}
(H_{4}\wedge\star H_{4})
+{1\over4}({\bf B}_{2}^{T}+
{\cal{A}}_{1}{\bf B}_{1}^{T}){\eta}{\bf H}_{3}\wedge\bigg[
2dB_{3}+{1\over2}({\bf B}_{2}^{T}+{\cal{A}}_{1}{\bf B}_{1}^{T}){\eta}
{\bf H}_{2} \cr
-{1\over2}{\bf B}_{1}^{T}{\eta}{\bf H}_{3}\bigg]
-{1\over2}dB_{3}{\wedge}dB_{3}{\wedge}{\cal{A}}_{1}.
\eea
%
Hence $Sl(2,\bb{R})$ generalised dimensional reduction of the bosonic
sector of IIB supergravity to D=9 leads to the following bosonic
lagrangian
\bea \label{so2massive}
e^{-1}{\cal{L}}_{9} &=& R -{1\over2}(\partial\varphi)^{2} + {1\over4}
tr(D{\cal{M}}D{\cal{M}}^{-1}) - {1\over4}e^{-{4\over\sqrt7}\varphi}
{\cal{F}}_{2}^{2} 
\nonumber \\
& &
-{1\over4}e^{{3\over\sqrt{7}}\varphi}
{\bf{\cal{H}}}_{2}^{T}{\cal{M}}{\bf{\cal{H}}}_{2}
-{1\over12}e^{-{1\over\sqrt7}\varphi}{\bf{\cal{H}}}_{3}^{T}{\cal{M}}
{\bf{\cal{H}}}_{3}
-{1\over48}e^{{2\over\sqrt{7}}\varphi}({\cal{H}}_{4})^{2}
\nonumber\\
& &
-{1\over2}e^{{4\over{\sqrt7}}\varphi}\;tr\bigg(
C^{2}+C{\cal{M}}^{-1}C^{T}{\cal{M}}\bigg)
+ e^{-1}{\cal{L}}_{cs} 
\eea
where ${\cal{L}}_{cs}$ can be read off from (\ref{Lcs}). The various
field strengths are defined by
\bea \label{H4}
{\cal{F}}_{2} &=& d{\cal{A}}_{1}, \nonumber\\
{\bf H}_{2}(x) &=& d{\bf B}_{1}-C{\bf B}_{2}, \nonumber\\
{\bf H}_{3}(x) &=& d{\bf B}_{2}-{\bf H}_{2}\wedge{\cal{A}}_{1}, \nonumber\\
D{\cal{M}} &=& {\partial}{\cal{M}}-(C^{T}{\cal{M}}+{\cal{M}}C)
{\cal{A}}_{1}, \nonumber\\
D{\cal{M}}^{-1} &=& {\partial}{\cal{M}}^{-1}+(C^{T}{\cal{M}}^{-1}
+{\cal{M}}^{-1}C^{T}){\cal{A}}_{1} \nonumber\\
H_{4} &=& dB_{3}-{1\over2}{\bf B}_{1}^{T}{\eta}{\bf H}_{3}
+{1\over2}({\bf B}_{2}^{T}+
{\cal{A}}_{1}{\bf B}_{1}^{T}){\eta}{\bf H}_{2}
\eea
\be
{\cal{M}}=e^{\phi}{\pmatrix {|{\tau}|^{2} & {\chi} \cr {\chi} & 1}}, \qquad
\eta = {\pmatrix{{}0 & 1 \cr -1 & 0}}, \qquad
\tau=\chi+ie^{-\phi}
\ee
\be
C ={1\over2}{\pmatrix{m_{1} & m_{2}+m_{3} \cr  m_{2}-m_{3} & -m_{1}}}
\ee
and where ${\cal{H}}_{4}\equiv H_{\mu\nu\rho\sigma}$ etc. The action is invariant under global scalings of the fields and under
global $Sl(2,\bb{R})$ transformations \cite{MO}. The action of the $Sl(2,\bb{R})$
transformations is
\be
{\cal{M}}^{\prime}={\Lambda}{\cal{M}}{\Lambda}^{T}, \quad
{\bf B}_{1}^{\prime}=({\Lambda}^{T})^{-1}{\bf B}_{1}, \quad
{\bf B}_{2}^{\prime}=({\Lambda}^{T})^{-1}{\bf B}_{2}.
\ee
As $C$ is an element of the Lie algebra of $Sl(2,\bb{R})$ it must transform
in the adjoint representation
\be
C^{\prime}=({\Lambda}^{T})^{-1}C{\Lambda}^{T}.
\ee
The action also has a local symmetry. Consider general coordinate
transformations in D=10 for which
\be
{\delta}{\hat x}^{\hat{\mu}}={\delta}^{{\hat{\mu}}z}{\rho}(x)
\label{gct}
\ee
{\it i.e.} the compactification coordinate $z$ transforms by an arbitrary
function of the remaining nine dimensional coordinates $x$.
In ordinary Kaluza-Klein reduction, this D=10 general coordinate
transformation becomes a gauge transformation of the Kaluza-Klein vector
after reduction to D=9,
\be
{\delta}{\cal{A}}_{\mu}(x)={\partial}_{\mu}{\rho}(x).
\ee
None of the other D=9 fields are charged under this $U(1)$ which is thus
a trivial gauge symmetry which always appears upon dimensional reduction
by one dimension (reduction by $n$-dimensions gives a trivial $U(1)^{n}$
gauge symmetry).

In generalised dimensional reductions, because one now allows the various
fields to have a dependence on the compactification coordinate $z$, a
higher dimensional general coordinate transformation of the form
(\ref{gct}) will induce transformations of fields other than the
Kaluza-Klein vector in the lower dimension, and thus the $U(1)$ gauge symmetry
is now non-trivial. This is certainly what happens in the case of
$Sl(2,\bb{R})$ generalised dimensional reduction of IIB to D=9.
Thus (\ref{so2massive}) is invariant under the local transformations of
the fields
\bea
{\cal{A}}_{1}&{\rightarrow}& {\cal{A}}_{1}+{\partial}\rho \nonumber\\
{\bf B}_{1}&{\rightarrow}&e^{-C{\rho}}{\bf B}_{1}\nonumber\\
{\bf B}_{2}&{\rightarrow}&e^{-C{\rho}}({\bf B}_{2}-d{\rho}{\bf B}_{1})
\nonumber\\
{\cal{M}}&{\rightarrow}&e^{{C^{T}}{\rho}}{\cal{M}}e^{C{\rho}}
\eea
or, in terms of field strengths
\bea
{\bf H}_{2}&{\rightarrow}&e^{-C{\rho}}{\bf H}_{2}\nonumber\\
{\bf H}_{3}&{\rightarrow}&e^{-C{\rho}}{\bf H}_{3}\nonumber\\
D{\cal{M}}&{\rightarrow}&e^{{C^{T}}{\rho}}(D{\cal{M}})e^{C{\rho}}\nonumber\\
D{\cal{M}}^{-1}&{\rightarrow}&e^{-C{\rho}}(D{\cal{M}}^{-1})e^{-{C^{T}}{\rho}}
\eea
with all other fields invariant. The Chern-Simons term changes by a total
derivative.
\section{The $m_{1}=m_{2}=0$ Limit}

In the previous section we used an $Sl(2,\bb{R})$ Scherk-Schwarz reduction
to obtain the bosonic sector of a maximally supersymmetric D=9
supergravity containing three mass parameters and an $SO(2)$ gauging.
In this paper we are primarily concerned with just the $SO(2)$ gauged
model with as few non-zero mass parameters as possible. In this section we will show how the latter can be obtained as
a truncation of the former by making the convenient choice $m_{1}=m_{2}=0$ and $m_{3}\equiv m$.

Setting $m_{1}=m_{2}=m_{3}=0$ in the action (\ref{so2massive}) one
recovers the maximal ungauged D=9 supergravity \cite{KK}. This corresponds
to performing an ordinary dimensional reduction from D=10. It follows
that it is consistent to choose just $m_{3} \equiv m$ to be non-vanishing.

We argued in the introduction that gauging maximal supergravity in D=9
requires certain potentials to be massive and this could only be achieved
consistently if a doublet of vectors is eaten in a Higg's mechanism.
As mentioned in section (2.2) the generalised reduction leads to a
modification in the field strength of ${\bf B}_{1}$. This allows one to
make a gauge transformation of ${\bf B}_{2}$ thus eliminating
${\bf B}_{1}$. The vector potential ${\bf B}_{1}$ also appears in the
Chern-Simons modifications to the field strength $H_{4}$ of the 3-form
$B_{3}$. Hence we will need to redefine $B_{3}$ in order to eliminate
${\bf B}_{1}$ completely from the action. We begin with the gauge transformation of
${\bf B}_{2}$,
\be
{\bf B}_{2}\ {\rightarrow}\ {\bf B}_{2}+ C^{-1}d{\bf B}_{1}
\label{higgsmech}
\ee
where $C^{-1}$ is also a constant matrix given by
\be
C^{-1}=-{2\over{\Delta}}{\pmatrix{m_{1} & m_{2}+m_{3} \cr
m_{2}-m_{3} & -m_{1}}}, \qquad {\Delta}={m_{3}}^{2}-{m_{2}}^{2}-{m_{1}}^{2}.
\ee
The effect of this transformation on ${\bf H}_{2}$ and ${\bf H}_{3}$
is
\bea
{\bf H}_{2}&{\rightarrow}& -C{\bf B}_{2} \nonumber \\
{\bf H}_{3}&{\rightarrow}& d{\bf B}_{2}+C{\bf B}_{2}\wedge{\cal{A}}_{1}
\equiv D{\bf B}_{2}.
\eea
The field strength of the 3-form $B_{3}$ was given in (\ref{H4}). Expanding
out this expression, $H_{4}$ can alternatively be given as
\be
H_{4} = dB_{3}-{1\over2}{\bf B}_{2}^{T}{\eta}C{\bf B}_{2}
+{1\over2}{\bf B}_{2}^{T}{\eta}d{\bf B}_{1}-
{1\over2}{\bf B}_{1}^{T}{\eta}d{\bf B}_{2}.
\ee
After making the gauge transformation (\ref{higgsmech}), $H_{4}$ becomes
\be
H_{4} = dB_{3}+{1\over2}d({\bf B}_{1}^{T}{\eta}{\bf B}_{2})
-{1\over2}{\bf B}_{2}^{T}{\eta}C{\bf B}_{2}.
\ee
Hence ${\bf B}_{1}$ can be eliminated if we redefine the field $B_{3}$ as
\be
B_{3}\ {\rightarrow}\ B_{3}-{1\over2}{\bf B}_{1}^{T}{\eta}{\bf B}_{2}.
\label{B3}
\ee
The only other terms in the action (\ref{so2massive}) where the vector
potential ${\bf B}_{1}$ appears is in the Chern-Simons term. It turns
out that the redefinitions (\ref{higgsmech}) and (\ref{B3}) are sufficient to
ensure the complete elimination of ${\bf B}_{1}$ from the action. After
discarding some total derivative terms the Chern-Simons terms can be written 
\be
-{1\over2}dB_{3}{\wedge}\bigg[dB_{3}{\cal{A}}_{1}-
{\bf B}_{2}^{T}{\eta}D{\bf B}_{2}\bigg]
-{1\over8}({\bf B}_{2}^{T}{\eta}C{\bf B}_{2}){\wedge}
({\bf B}_{2}^{T}{\eta}D{\bf B}_{2}).
\ee
The following lagrangian is now in a form from which one can take the
$m_{1}=m_{2}=0$ limit \footnote{One should use (\ref{so2massive}) for the $C=0$ limit
and not (\ref{intermediate}). One could perform the transformation (\ref{higgsmech}) after having
set $m_{1}=m_{2}=0$ in (\ref{so2massive}).}.
\newpage
\bea
\label{intermediate}
I_{9}={\int}d^{9}x\,e\!\!\!\!\!\!\!\!&&\bigg\{
R-{1\over2}(\partial\varphi)^{2} + {1\over4}
tr(D{\cal{M}}D{\cal{M}}^{-1}) - {1\over4}e^{-{4\over\sqrt7}\varphi}
{\cal{F}}_{2}^{2} 
\nonumber \\
&&
-{1\over12}e^{-{1\over\sqrt7}\varphi}(D{\bf{\cal{B}}}_{2})^{T}{\cal{M}}
(D{\bf{\cal{B}}}_{2})
-{1\over4}e^{{3\over\sqrt{7}}\varphi}
{\bf{\cal{B}}}_{2}^{T}(C^{T}{\cal{M}}C){\bf{\cal{B}}}_{2}
\nonumber \\
&&
-{1\over48}e^{{2\over\sqrt{7}}\varphi}({\cal{H}}_{4})^{2}
-{1\over2}e^{{4\over{\sqrt7}}\varphi}\;tr\bigg(
C^{2}+C{\cal{M}}^{-1}C^{T}{\cal{M}}\bigg)\bigg\}
\nonumber \\
&&
\!\!\!\!\!\!\!\!\!\!\!\!\!\!\!\!\!\!\!\!\!\!\!
+\int\,-{1\over2}\,dB_{3}{\wedge}\bigg(dB_{3}{\cal{A}}_{1}-
{\bf B}_{2}^{T}{\eta}D{\bf B}_{2}\bigg)
-{1\over8}({\bf B}_{2}^{T}{\eta}C{\bf B}_{2}){\wedge}
({\bf B}_{2}^{T}{\eta}D{\bf B}_{2}).
\eea
where
\bea
{\cal{F}}_{2} &=& d{\cal{A}}_{1}, \nonumber\\
D{\bf B}_{2} &=& d{\bf B}_{2}+C{\bf B}_{2}{\cal{A}}_{1}, \nonumber\\
H_{4} &=& dB_{3}-{1\over2}{\bf B}_{2}^{T}{\eta}C{\bf B}_{2}, \nonumber\\
D{\cal{M}} &=& {\partial}{\cal{M}}-(C^{T}{\cal{M}}+{\cal{M}}C)
{\cal{A}}_{1}, \nonumber\\
D{\cal{M}}^{-1} &=& {\partial}{\cal{M}}^{-1}+(C^{T}{\cal{M}}^{-1}
+{\cal{M}}^{-1}C^{T}){\cal{A}}_{1},
\eea
\be
{\cal{M}}=e^{\phi}{\pmatrix {|{\tau}|^{2} & {\chi} \cr {\chi} & 1}},
\qquad
\eta = {\pmatrix{{}0 & 1 \cr -1 & 0}}, \qquad
\ee
\be
C ={1\over2}{\pmatrix{m_{1} & m_{2}+m_{3} \cr  m_{2}-m_{3} & -m_{1}}}=
-{{\Delta}\over4}C^{-1},
\ee
\be
\tau=\chi+ie^{-\phi}, \qquad
{\Delta}={m_{3}}^{2}-{m_{2}}^{2}-{m_{1}}^{2}
\label{fielddefs1}
\ee
and where ${\cal{H}}_{4}\equiv H_{\mu\nu\rho\sigma}$ etc.
\subsection{N=2 D=9 $SO(2)$ Gauged Supergravity}

We can now set $m_{1}=m_{2}=0$ and $m_{3}\equiv m$. There are no problems
with the matrix $C$ or its inverse in this limit which simply becomes
$C={m\over2}{\eta}, \ C^{-1}=-{2\over m}{\eta}$. It is interesting to
examine the scalar potential in this limit
\be
{\cal{V}}({\cal{M}})={1\over2}\,e^{{4\over{\sqrt7}}\varphi}\;tr\bigg(
C^{2}+C{\cal{M}}^{-1}C^{T}{\cal{M}}\bigg)
\ee
becomes
\be
{\cal{V}}({\cal{M}})={{m^{2}}\over8}\,e^{{4\over{\sqrt7}}\varphi}
\;tr\bigg(-{{\bf 1}_{2}}+{\eta}{\cal{M}}^{-1}{\eta}^{T}{\cal{M}}\bigg)
={{m^{2}}\over8}\,e^{{4\over{\sqrt7}}\varphi}
\;tr\bigg(-{{\bf 1}_{2}}+{\cal{M}}^{2}\bigg).
\ee
 Making the above replacements in (\ref{intermediate}) and using again
the $Sl(2,\bb{R})$ property
${\cal{M}}^{-1}={\eta}^{T}{\cal{M}}{\eta}$ we finally get the bosonic
sector of a maximally extended $SO(2)$ gauged supergravity in nine
dimensions:
\bea \label{so2gauged}
I_{9}^{SO(2)}={\int}d^{9}x\,e\!\!\!\!\!\!\!\!
&&\bigg\{R -{1\over2}(\partial\varphi)^{2}+{1\over4}
tr(D{\cal{M}}D{\cal{M}}^{-1})-{1\over4}e^{-{4\over\sqrt7}\varphi}
{\cal{F}}_{2}^{2} 
\nonumber \\
& &
-{1\over12}e^{-{1\over\sqrt7}\varphi}(D{\bf{\cal{B}}}_{2})^{T}{\cal{M}}
(D{\bf{\cal{B}}}_{2})
-{{m}^{2}\over16}e^{{3\over\sqrt{7}}\varphi}
{\bf{\cal{B}}}_{2}^{T}{\cal{M}}^{-1}{\bf{\cal{B}}}_{2}
\nonumber \\
& &
-{1\over48}e^{{2\over\sqrt{7}}\varphi}({\cal{H}}_{4})^{2}
-{{m}^{2}\over8}e^{{4\over{\sqrt7}}\varphi}\;tr\bigg(
{\cal{M}}^{2}-{\bf 1}_{2}\bigg)\bigg\}
\nonumber \\
& &
\!\!\!\!\!\!\!\!\!\!\!\!\!\!\!\!\!\!\!\!\!\!\!
-{1\over2}\int\,dB_{3}{\wedge}\bigg(dB_{3}{\cal{A}}_{1}-
{\bf B}_{2}^{T}{\eta}D{\bf B}_{2}\bigg)
-{m\over8}({\bf B}_{2}^{T}{\bf B}_{2}){\wedge}
({\bf B}_{2}^{T}{\eta}D{\bf B}_{2}).
\eea
where
\bea \label{DM}
{\cal{F}}_{2} &=& d{\cal{A}}_{1}, \nonumber\\
D{\bf B}_{2} &=& d{\bf B}_{2}+{m\over2}{\eta}{\bf B}_{2}{\cal{A}}_{1}, \nonumber\\
H_{4} &=& dB_{3}+{m\over4}{\bf B}_{2}^{T}{\bf B}_{2},
\nonumber\\
D{\cal{M}} &=& {\partial}{\cal{M}}-{m\over2}({\eta}^{T}{\cal{M}}+{\cal{M}}\eta)
{\cal{A}}_{1}, \nonumber\\
D{\cal{M}}^{-1} &=& {\partial}{\cal{M}}^{-1}+{m\over2}({\eta}^{T}{\cal{M}}^{-1}
+{\cal{M}}^{-1}{\eta}^{T}){\cal{A}}_{1},
\eea
\be \label{Mtau}
{\cal{M}}=e^{\phi}{\pmatrix {|{\tau}|^{2} & {\chi} \cr {\chi} & 1}}, 
\qquad
\eta = {\pmatrix{{}0 & 1 \cr -1 & 0}}, \qquad
\tau=\chi+ie^{-\phi}
\ee
and where ${\cal{H}}_{4}\equiv H_{\mu\nu\rho\sigma}$ etc.
The action (\ref{so2gauged}) is invariant under the following local
$SO(2)$ transformations
\bea
{\cal{A}}_{1} & {\rightarrow} & {\cal{A}}_{1}+{\partial}\rho \nonumber\\
{\bf B}_{2} & {\rightarrow} & e^{-{m\over2}\eta{\rho}}{\bf B}_{2}
\nonumber\\
{\cal{M}} & {\rightarrow} & e^{-{m\over2}\eta{\rho}}{\cal{M}}
e^{{m\over2}\eta{\rho}}.
\eea
In terms of field strengths
\bea
D{\bf B}_{2} & {\rightarrow} & e^{-{m\over2}\eta{\rho}}D{\bf B}_{2}
\nonumber\\
D{\cal{M}} & {\rightarrow} & e^{-{m\over2}\eta{\rho}}(D{\cal{M}})
e^{{m\over2}\eta{\rho}} 
\nonumber\\
D{\cal{M}}^{-1} & {\rightarrow} & e^{-{m\over2}\eta{\rho}}(D{\cal{M}}^{-1})
e^{{m\over2}\eta{\rho}}.
\eea
Clearly the Chern-Simons term is invariant up to a total
derivative.
\section{D=9 Gauged Supergravity Vacua}

In this section we shall look for 7-brane domain wall solutions of the D=9
SO(2) gauged model (\ref{so2gauged}). Examination of the field equations of this model shows
that one can consistently truncate the model to one containing just three
scalars (and gravity). We are interested in 7-brane domain walls so the metric we choose is
\be \label{domwall}
dS_{9}^{2} = e^{2A(r)}dx^{\mu}dx^{\nu}{\eta}_{\mu\nu}+dr^{2}
\qquad {\mu,\nu}=0,\dots,7.
\ee
Note that this metric becomes $Mink_{9}$ for $A(r)$ constant. The field
equations then become (primes denote differentiation w.r.t. r.)
\be \label{fieldeqn1}
{\varphi}^{\prime\prime}+8{A}^{\prime}{\varphi}^{\prime}=
{{m^{2}}\over{2\sqrt7}}\;e^{{4\over{\sqrt7}}\varphi}[e^{2\phi}({\chi}^{2}+1)^{2}+e^{-2\phi}+2({\chi}^{2}-1)]
\ee
\be
{\phi}^{\prime\prime}+8{A}^{\prime}{\phi}^{\prime}-e^{2\phi}({\chi}^{\prime})^{2}={{m^{2}}\over{4}}e^{{4\over{\sqrt7}}\varphi}[e^{2\phi}({\chi}^{2}+1)^{2}-
e^{-2\phi}]
\ee
\be \label{chieqn}
{\chi}^{\prime\prime}+2{\phi}^{\prime}{\chi}^{\prime}+
8{A}^{\prime}{\chi}^{\prime}={{m^{2}}\over{2}}{\chi}\;e^{{4\over{\sqrt7}}\varphi}
[({\chi}^{2}+1)+e^{-2\phi}]
\ee
\be \label{Aeqn}
{\phi}^{\prime}{\chi}={1\over2}{\chi}^{\prime}[e^{2\phi}({\chi}^{2}+1)-
1]
\ee
\be
28({A}^{\prime})^{2}={1\over4}[({\varphi}^{\prime})^{2}+({\phi}^{\prime})^{2}
+e^{2\phi
}({\chi}^{\prime})^{2}]-{{m^{2}}\over{16}}\;e^{{4\over{\sqrt7}}\varphi}
[e^{2\phi}({\chi}^{2}+1)^{2}+e^{-2\phi}+2({\chi}^{2}-1)]
\ee
\be \label{fieldeqn2}
7({A}^{\prime\prime}+4({A}^{\prime})^{2})=-{1\over4}[({\varphi}^{\prime})^{2}+({\phi}^{\prime})^{2}+e^{2\phi}({\chi}^{\prime})^{2}]
-{{m^{2}}\over{16}}\;e^{{4\over{\sqrt7}}\varphi}
[e^{2\phi}({\chi}^{2}+1)^{2}+e^{-2\phi}+2({\chi}^{2}-1)]
\ee
The rather curious equation (\ref{Aeqn}) has its origin in the field equation
for the gauge potential ${\cal{A}}_{\mu}$. The remaining field equations can be derived from the lagrangian
\be \label{lagtrunc}
e^{-1}{\cal{L}} = R -{1\over2}(\partial\varphi)^{2}
-{1\over2}(\partial\phi)^{2}-{1\over2}{e^{2\phi}}(\partial\chi)^{2}
-{\cal{V}}({\varphi,\phi,\chi})
\ee
where
\be
{\cal{V}}({\varphi,\phi,\chi}) = {{m^{2}}\over8}e^{{4\over{\sqrt7}}\varphi}[e^{2\phi}({\chi}^{2}+1)^{2}+e^{-2\phi}+2({\chi}^{2}-1)].
\ee
The Bogomol'nyi equations for the system are \cite{GibLamb,ChamGib}
\be
{A}^{\prime}={\mp}2W, \qquad {\Phi}^{\prime{A}}={\pm}(14){\gamma}^{AB}
{{{\partial}W}\over{\partial{\Phi}^{B}}}
\ee
where ${\Phi}^{T}=(\varphi\;\;\phi\;\;\chi)$ and ${\gamma}_{AB}=
{1\over2}diag(1,1,e^{2\phi})$. The superpotential $W$ is related to the scalar
potential via \cite{GibLamb}
\be
{\cal{V}}=4(D-2)^{2}[{\gamma}^{AB}{{{\partial}W}\over{\partial{\Phi}^{A}}}
{{{\partial}W}\over{\partial{\Phi}^{B}}}-{(D-1)\over{(D-2)}}W^{2}]
\ee
where $D$ is the spacetime dimension. Solving this equation leads to 
\be \label{superpot}
W(\varphi,\phi,\chi)={{m}\over{56}}\;e^{{2\over{\sqrt7}}\varphi}
[e^{\phi}({\chi}^{2}+1)+e^{-\phi}+2c]
\ee
where $c$ is an arbitrary constant. This constant arises because the term in
${\cal{V}}$ proportional to $2({{{\partial}W}\over{\partial{\varphi}}})^{2}$
is exactly equal to ${(D-1)\over{(D-2)}}W^{2}$ in $D=9$. Thus the Bogomol'nyi
equations for the domain wall background are 
\be \label{BogomolA}
{\varphi}^{\prime}={\pm}{m\over{\sqrt7}}\;e^{{2\over{\sqrt7}}\varphi}
[e^{\phi}({\chi}^{2}+1)+e^{-\phi}+2c]
\ee
\be
{\phi}^{\prime}={\pm}{m\over{2}}\;e^{{2\over{\sqrt7}}\varphi}
[e^{\phi}({\chi}^{2}+1)-e^{-\phi}]
\ee
\be
{\chi}^{\prime}={\pm}m{\chi}e^{{2\over{\sqrt7}}\varphi}e^{-\phi}
\ee
\be \label{BogomolB}
{\varphi}^{\prime}=-4{\sqrt7}{A}^{\prime}.
\ee
We have shown these equations imply the field equations (\ref{fieldeqn1}$\,$-$\,$\ref{fieldeqn2}). The field equations and Bogomol'nyi equations can be systematically investigated by considering
the eight possible cases for which each of the three scalars are independently
turned on or off. Of these cases the field equations are only potentially
soluble in four cases (for example, the choice $\phi=0$ with $\varphi$ and
$\chi$ non-zero solves (\ref{Aeqn}) if $\chi$ is constant but one cannot then solve the $\chi$ equation of motion (\ref{chieqn})).

\vs{10}
\noindent Case i) \ $\varphi=\chi=\phi=0$

The field equations are trivially solved with $A(r)$ constant. This is 9D Minkowski spacetime, a somewhat surprising solution of a gauged maximal
supergravity. Examination of the Bogomol'nyi equations show this is only a solution if $c=-1$, in which case ${{{\partial}W}\over{\partial{\varphi}}}=
{{{\partial}W}\over{\partial{\phi}}}={{{\partial}W}\over{\partial{\chi}}}=0$
and $Mink_9$ is a maximally supersymmetric solution.

\vs{10}
\noindent Case ii) \ $\varphi\neq 0,\;\chi=\phi=0$

The field equations are all identically satisfied except the $\varphi$ and ${\cal{A_{\mu}}}$ equations and Einstein's equation which become
\be \label{topeqn}
{\varphi}^{\prime\prime}+8{A}^{\prime}{\varphi}^{\prime}=0
\ee
\be \label{middleeqn}
({A}^{\prime})^{2}={1\over{(4)(28)}}({\varphi}^{\prime})^{2}
\ee
\be \label{lasteqn}
{A}^{\prime\prime}+4({A}^{\prime})^{2}=-{1\over28}({\varphi}^{\prime})^{2}.
\ee
Solving (\ref{middleeqn}) and substituting in (\ref{topeqn}) and (\ref{lasteqn}) leads to the single
equation for $\varphi$
\be
{\varphi}^{\prime\prime}{\pm}{2\over{\sqrt7}}({\varphi}^{\prime})^{2}=0
\ee
and the metric function can be found from $A(r)={\pm}{1\over{4\sqrt7}}\varphi(r)+\alpha$. The
constant $\alpha$ can be dropped as this corresponds to a simple rescaling
of the coordinates. One solution of these equations is $\varphi(r)$ and
$A(r)$ both constant {\it i.e.} $Mink_9$ with an arbitrary constant scalar
$\varphi$. This does not solve the Bogomol'nyi equations unless
$c=-1$ and is a generalisation of case i) above. Another interesting solution of the field equations is the domain wall
with the single scalar $\varphi$ active
\be
e^{\pm{2\over{\sqrt7}}\varphi}=H(r)=ar+b
\ee
\be
dS_{9}^{2} = H^{1\over4}(r)dx^{\mu}dx^{\nu}{\eta}_{\mu\nu}+dr^{2}
\qquad {\mu,\nu}=0,\dots,7.
\ee
where $a$ and $b$ are arbitrary constants. However, this domain wall only
becomes a solution of the Bogomol'nyi equations if $a$ is proportional to
$c+1$. Thus in the the purely bosonic theory for which $c$ is not fixed
this domain wall satisfies the Bogomol'nyi equations assuming $c\neq -1$ and is thus stable.
It will turn out though that $c=-1$ in the supergravity model and thus
$H(r)$ is constant and the solution reduces to maximally supersymmetric
$Mink_9$ with an arbitrary constant scalar.

\vs{10}
\noindent Case iii) \ $\chi=0,\;\varphi,\phi\neq 0$

\begin{figure}
\psfrag{G}{$G(\phi)$}
\psfrag{f}{$\phi$}
\begin{center}
\includegraphics{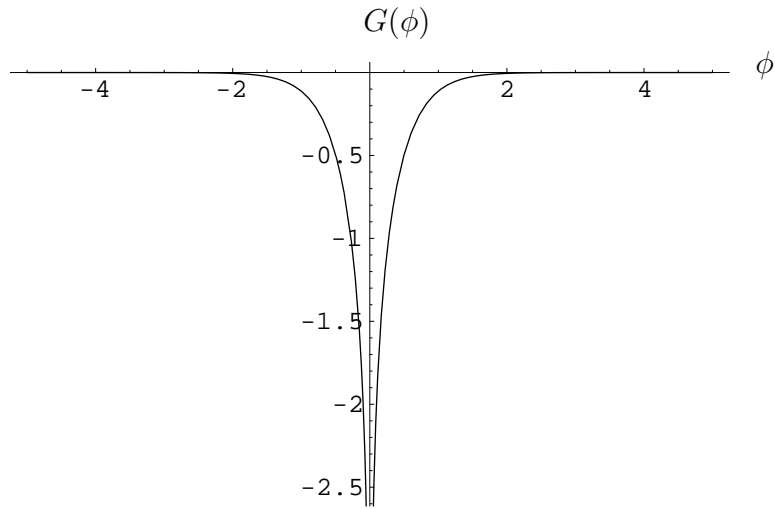}
\end{center}
\caption{\label{fig:Wall}\small{Functional dependence of $G(\phi)$ on $\phi$.}}
\end{figure}

Here we can try and directly solve the Bogomol'nyi equations which are
\be \label{topeqna}
{\varphi}^{\prime}={\pm}{m\over{\sqrt7}}\;e^{{2\over{\sqrt7}}\varphi}
[e^{\phi}+e^{-\phi}+2c]
\ee
\be \label{middleeqna}
{\phi}^{\prime}={\pm}{m\over{2}}\;e^{{2\over{\sqrt7}}\varphi}
[e^{\phi}-e^{-\phi}]
\ee
\be \label{lasteqna}
{\varphi}^{\prime}=-4{\sqrt7}{A}^{\prime}.
\ee
Clearly the metric function $A(r)$ can be easily found from (\ref{lasteqna}) once $\varphi(r)$ is known. We therefore concentrate on the first two equations above. We
lose no generality by working with the upper (positive) signs as $m$ is arbitrary. Substitution of (\ref{topeqna}) into (\ref{middleeqna}) leads to
\be
{\mu}e^{{\sqrt{7}\over2}\varphi}=\left\vert{{(\sinh({\phi\over{2}}))^{c+1}}\over
{(\cosh({\phi\over{2}}))^{c-1}}}\right\vert
\ee
where $\mu$ is an arbitrary positive constant. On further substitution of
this equation into (\ref{middleeqna}) one obtains the following integral
\be
\int d\phi \ \frac{({\cosh\phi})^{({4c-11})\over7}}{({\sinh\phi})^{({4c+11})\over7}} =H(r)=\alpha mr+\beta
\ee
where $\alpha$ is an arbitrary positive constant and $\beta$ is just
an arbitrary constant. For the supersymmetric case, $c=-1$,
we have the equations
\be
e^{{\sqrt{7}\over2}\varphi(r)}=\cosh^{2}\left({{\phi(r)}\over{2}}\right)
\ee
\be \label{Gphieqn}
{7\over8{(\cosh\phi)^{8\over7}}}+\int d\phi\ 
{1\over{(\sinh\phi)(\cosh\phi)^{1\over7}}}=H(r)=2mr+\beta
\ee
where $\mu$ has been set to one. After making a change of variables, the integral on the left hand side of (\ref{Gphieqn}) can be expressed in terms of the hypergeometric function ${}_{2}{F}_{1}$ defined as follows
\be
{ }_{2}F_{1}(a,b,c\; ;z)={{\Gamma(c)}\over{\Gamma(b)\Gamma(c-b)}}{\int}_{0}^{1}
dt \ t^{b-1}(1-t)^{c-b-1}(1-tz)^{-a}.
\ee
Thus the dependence of $\phi$ on the transverse distance $r$ from the domain wall is described by the function $G(\phi)$ 
\be
G(\phi)\equiv {7\over8}\Bigg\{{1\over{(\cosh\phi)^{8\over7}}}-
{1\over{(\sinh\phi)^{8\over7}}}\ { }_{2}F_{1}({4\over7},{4\over7},{11\over7},
-cosech^{2}\phi)\Bigg\}=H(r).
\ee
One can easily show $G(0)=-\infty$ and for $|\phi|\rightarrow\infty,\ 
{ }_{2}F_{1}({4\over7},{4\over7},{11\over7},-cosech^{2}\phi)\rightarrow 1$ and hence $G(\phi)\rightarrow 0$ from below. $G^{\prime}(\phi)=0$ only at $|\phi|=\infty$ thus a plot of the function $G(\phi)$ versus $\phi$ takes the form
of figure \ref{fig:Wall}.
This plot shows $G(\phi)$ and hence $H(r)$ are negative for all values of $\phi$. 

In a typical domain wall solution ({\it{e.g.}} IIA 8-brane) the scalar field
$\phi$ is related to an harmonic function $H(r)$ via $e^{\kappa\phi}=H(r)$ and hence $H(r)$ is always positive although $H^{\prime}(r)$ can take either sign. Choosing $H(r)=\alpha r+\beta$ and imposing that $H(r)$ is positive leads to two inequalities for $r$ depending on the sign of $\alpha$, $r>-{\beta\over{|\alpha|}}$ for $\alpha>0$ and $r<{\beta\over{|\alpha|}}$ for $\alpha<0$. One can then choose $\beta$ differently in these two regions to make $H(r)$ continuous {\it{i.e.}} $\beta=-|\alpha|{r}_{0}$ where $\alpha>0$ and $\beta=|\alpha|{r}_{0}$ for $\alpha<0$. The continuous function $H(r)$ is then positive for all $r$ and $H^{\prime}(r)$ changes sign in going from $r<{r}_{0}$ to $r>{r}_{0}$. Thus $H(r)=|\alpha||r-{r}_{0}| \ \ \forall \; r$.

The difference of the discussion of the previous paragraph and our case is that the Bogomol'nyi equations imply that the coefficient of $r$ in $H(r)$ must always be positive, assuming $m>0$ (or at least has the same sign everywhere).
Combined with the requirement of $H(r)$ always being negative, this leads to just one inequality for $r$, $H(r)=2mr+\beta$ is negative for $r<-{\beta\over{2m}}$. Hence we have a single sided domain wall.

One can avoid this situation if one is prepared to allow $m$ to be positive in one region and negative in another. Then demanding $H(r)<0 \ \ \forall \; r$ leads again to two inequalities and with $\beta$ chosen appropriately in these two regions as above we can solve the Bogomol'nyi equations everywhere with
\bea \label{harmfunca}
H(r)=-2|m|(r-{r}_{0}) \qquad r>{r}_{0}
\nonumber\\
H(r)=+2|m|(r-{r}_{0}) \qquad r<{r}_{0}
\eea
{\it{i.e.}} $H(r)=2m(r-{r}_{0})$ where $m=\mp |m|$ for $r>{r}_{0}$ and $r<{r}_{0}$ respectively. 

Examination of the Bogomol'nyi equations show that allowing $m$ to change sign in different regions of the spacetime is equivalent to keeping $m$ fixed but working with the Bogomol'nyi equations with the opposite sign choice in different regions but which still imply the same field equations. This procedure is justified by the fact that $m$ appears in the field equations (\ref{fieldeqn1}$\,$-$\,$\ref{fieldeqn2}) and Lagrangian
(\ref{lagtrunc}) quadratically so one can consider only $|m|$ to be fixed. However $m$ does appear linearly in covariant derivatives but these terms vanish after truncating to the bosonic sector of scalars and gravity.

The harmonic function (\ref{harmfunca}) tends to $-\infty$ as $|r|\rightarrow\infty$. Hence the scalars $\phi$ and $\varphi$ and the metric function $A$ all vanish as $|r|\rightarrow\infty$ making the geometry in this region $Mink_9$.
At $r=r_{0}$, $H(r)$ vanishes and hence $|\phi|=\varphi=\infty$ and $A=-\infty$. Using the expression for the superpotential $W(\varphi,\phi,\chi)$ with $c=-1$, (\ref{superpot}), we observe that the sign of $W$ is different on the two sides of the domain wall due to $m$ changing sign. $|W|\rightarrow 0$ as $|r|\rightarrow\infty$ and $|W|\rightarrow\infty$ as $|r|\rightarrow {r}_{0}$.

One can obtain a domain wall for which $\phi$ and $\varphi$ remain finite at $r={r}_{0}$ by taking $H(r)$ to be
\bea \label{harmfuncb}
H(r)=-2|m|(r-{r}_{0}+\lambda) \qquad r>{r}_{0}
\nonumber\\
H(r)=+2|m|(r-{r}_{0}-\lambda) \qquad r<{r}_{0}
\eea
where $\lambda>0$. We note that this is not the Randall-Sundrum scenario as at $r={r}_{0}$ where $\varphi$ is a positive constant, $A(r)$ is negative due to $\varphi^{\prime}=-4\sqrt{7}A^{\prime}$. Thus the gravitational potential $g_{00}$ increases from $r={r}_{0}$
to 
$|r|=\infty$ where $A=0$ and the geometry is Minkowski. Therefore this domain wall does not apparently confine the zero mode of the graviton to the brane nor is the geometry asymptotically Anti De-Sitter.

In summary, the supersymmetric domain wall is described by
\be
dS_{9}^{2} = e^{2A(r)}dx^{\mu}dx^{\nu}{\eta}_{\mu\nu}+dr^{2}
\qquad {\mu,\nu}=0,\dots,7
\ee
\be
G(\phi)\equiv {7\over8}\Bigg\{{1\over{(\cosh\phi)^{8\over7}}}-
{1\over{(\sinh\phi)^{8\over7}}}\ { }_{2}F_{1}({4\over7},{4\over7},{11\over7},
-cosech^{2}\phi)\Bigg\}=H(r),
\ee
\be
e^{-7A(r)}=e^{{\sqrt7\over4}\varphi(r)}=\cosh\left(\frac{\phi(r)}{2}\right),
\ee
where $H(r)$ is an harmonic function of $r$. This solution satisfies the Bogomol'nyi equations for $c=-1$ and therefore presumably preserves one half of the supersymmetry.

\vs{10}
\noindent Case iv) $\varphi,\phi,\chi \neq 0$

For this case we have been unable to find a solution of the full set of Bogomol'nyi equations (\ref{BogomolA}$\,$-$\,$\ref{BogomolB}) with all three scalars active.
%
\subsection{IIB Interpretation of 9D Vacua}

As the 9D SO(2) gauged supergravity (\ref{so2gauged}) was obtained by a generalised reduction
from D=10 type IIB supergravity, the 9D Minkowski spacetime and 7-brane domain wall solutions of the previous section should be the generalised
reductions of objects in D=10. We now address this point. In section 4 we took the constant $c$ appearing
in the 9D Bogomol'nyi equations to have the value $-1$ in the supersymmetric
case. We justify this here by the following outlined method.

One begins with the D=10 Bogomol'nyi equations in the 7-brane background \cite{GGP,GSVY}
\be
dS_{10}^{2} = dx^{\mu}dx^{\nu}{\eta}_{\mu\nu}+{\Omega}^{2}(dr^{2}+dx_{2}^{2})
 \qquad {\mu,\nu =0,\dots,7}
\ee
where ${\Omega}={\Omega}(r,x_{2})$, which are {\footnote{${\bar{\partial}}\equiv {1\over2}({\partial\over{\partial r}}+i{\partial\over{\partial x_{2}}})$ and ${\hat \tau}={\hat \tau}_{1}+i{\hat \tau}_{2}$. See \cite{GGP} for further details of the notation.}}
\be \label{10DBogomol}
{\bar{\partial}}{\hat \tau}=0, \qquad {\bar{\partial}}\ln \Omega =
{i\over{4{\hat{\tau}}_{2}}}\left({{\hat \tau}-i}\over{{\bar{\hat \tau}-i}}\right){\bar{\partial}}{\bar{\hat \tau}}
\ee
where ${\hat \tau}={\hat{\chi}}+ie^{-{\hat{\phi}}}$ is the D=10 complex scalar.
One then performs a generalised reduction of these Bogomol'nyi equations making use of the formula (derivable from the ansatz (\ref{genredansatz}))
\be \label{tauansatz}
{\hat \tau}(r,z) = {{{1\over2}(|\tau(r)|^{2}-1)\sin(mz)+{\chi(r)}\cos(mz)+ie^{-{\phi(r)}}}\over{(|\tau(r)|^{2}-1)\sin^{2}({mz\over{2}})+{\chi(r)}\sin(mz)+1}}
\ee
where $z$ is the compactification coordinate of section 2, related to $x_2$
by $dx_{2}=e^{-{2\over\sqrt{7}}\varphi(r)}dz$.
Comparison of the above metric with (\ref{metricansatz}) and (\ref{domwall})
immediately gives this relation between
$z$ and $x_{2}$ and the relation $\varphi (r)=-4\sqrt{7} A(r)$. After a very lengthy calculation, all $z$ dependence cancels and the real and imaginary parts of the (generalised)
reduced ${\bar{\partial}}{\hat \tau}=0$ equation give
\be \label{9DBogomola}
{\chi}^{\prime}={-}m{\chi}e^{{2\over{\sqrt7}}\varphi}e^{-\phi}
, \qquad {\phi}^{\prime}={-}{m\over{2}}\;e^{{2\over{\sqrt7}}\varphi}
[e^{\phi}({\chi}^{2}+1)-e^{-\phi}].
\ee
The imaginary part of the (generalised) reduction of the second equation of
(\ref{10DBogomol}) is identically satisfied whereas the real part becomes the equation
\be \label{9DBogomolb}
{\varphi}^{\prime}={m{\sqrt7}}\;e^{{2\over{\sqrt7}}\varphi}
[e^{\phi}({\chi}^{2}+1)+e^{-\phi}-2]
\ee
thus confirming that $c=-1$ in the supersymmetric case \footnote{This equation differs from (\ref{BogomolA}) by a factor of $-{1\over7}$. We believe this
disagreement originates in the differences of notion and conventions used in this paper compared to \cite{GGP,GSVY} and is not a serious one.}.

The domain wall of section 4 is a solution of the Bogomol'nyi equations (\ref{BogomolA})\ -\ (\ref{BogomolB})
with $\chi=0$ and therefore, due to the consistency of the Kaluza-Klein reduction, is a solution of the D=10 Bogomol'nyi equations in the 7-brane
background. Thus the generalised reduction of the type IIB D7-brane is the 9D
supersymmetric domain wall of section 4. The 9D Minkowski vacuum has an equally simple IIB interpretation. Setting $\varphi=\chi=\phi=A=0$ and $\tau=i$ in
(\ref{tauansatz})
implies ${\hat\tau}(x,z)=i$. Substituting into the D=10 Bogomol'nyi equations
gives a D=10 solution provided $\Omega(r,x_{2})$ is constant and hence $dS_{10}^{2}=Mink_{10}$. {\it{i.e.}} the IIB D9-brane. Thus the 9D Minkowski vacuum
is a wrapped D9-brane.

It is interesting that the $Mink_{10}$ solution of IIB supergravity survives the generalised reduction when one uses the global non-compact $Sl(2,\bb{R})$ symmetry as in this paper.
This is in contrast to the generalised reduction of IIB supergravity using a
different symmetry - the global non-compact shift symmetry of the axion
${\hat{\chi}}(x,z)$ {\footnote{This gives a 9-D massive supergravity which
is also the reduction of the massive IIA model \cite{BDRGPT}.}}. The generalised ansatz for
${\hat{\chi}}$ is ${\hat{\chi}}(x,z)=\chi(x)+mz$. Thus the $Mink_{10}$ solution, for which ${\hat{\chi}}=0$, is inconsistent with the ansatz and hence the
9D generalised reduced model has no $Mink_9$ solution.

\section{Further Dimensional Reductions}

The $SO(2)$ gauged supergravity (\ref{so2gauged}) has a Minkowski
ground state. An interesting question is whether, after ordinary dimensional
reduction to lower $D$, the resulting supergravities also admit
Minkowski vacua.

Upon performing standard Kaluza-Klein reduction to D=8 ({\it i.e.} all
fields are taken to be independent of the compactification coordinate),
it is clear that the potential simply reduces to
\be
{\cal{V}}({\cal{M}})={{m^{2}}\over8}\, e^{{4\over{\sqrt7}}\varphi}
\;tr\bigg( {\cal{M}}^{2}-{{\bf{1}}_{2}}\bigg)e^{2{\alpha}_{8}{\sigma}_{8}}
\ee
where the metric ans\"atz is taken to be
\be
dS_{9}^{2} = e^{2{\alpha}_{8}{\sigma}_{8}}dS_{8}^{2} + 
e^{2{\beta}_{8}{\sigma}_{8}}(dz+{\cal{A}}_{2}). 
\ee
However, because of the presence of a covariant derivative in the kinetic
term for ${\cal{M}}(\chi,\phi)$ there will be a contribution to the potential
from the term
\be
{1\over4}e\,tr(D{\cal{M}}D{\cal{M}}^{-1}).
\ee
Of course it was this scalar kinetic term which generated the potential
originally in the generalised reduction to D=9. Now it is the presence of
${\cal{A}}_{z}$ which is responsible for the new contribution to the
potential. Also if we reduce to low
enough dimension there will be new contributions to the potential from
the terms
\be \label{termsinB2}
-{1\over12}e\,e^{-{1\over\sqrt7}\varphi}(D{\bf{\cal{B}}}_{2})^{T}{\cal{M}}
(D{\bf{\cal{B}}}_{2})\qquad{\rm and}\qquad
-{{m}^{2}\over16}e^{{3\over\sqrt{7}}\varphi}
{\bf{\cal{B}}}_{2}^{T}{\cal{M}}^{-1}{\bf{\cal{B}}}_{2}.
\ee
We now analyse the structure of these contributions more closely.

Consider,
\be
-{1\over4}{\hat e}\;tr\bigg({D}_{{\hat{\mu}}}{\hat{\cal{M}}}
{D}^{{\hat{\mu}}}{\hat{\cal{M}}}^{-1}\bigg)
\ee
where ${D}_{{\hat{\mu}}}{\hat{\cal{M}}}$ and 
${D}^{{\hat{\mu}}}{\hat{\cal{M}}}^{-1}$ are given in (\ref{DM}) but with
${\cal{M}}$ replaced by ${\hat{\cal{M}}}$\footnote{Hats now refer to D=9
fields.}. For the reduction ans\"atze we use
\bea
{\hat{\cal{M}}}({\hat\chi}(x,z), {\hat\phi}(x,z)) &=& 
{\cal{M}}({\chi}(x), {\phi}(x)) \nonumber\\
{\hat{\cal{A}}}_{1}(x,z) &=& ({\cal{A}}_{1}-{\lambda}_{8}{\cal{A}}_{2})+
{\lambda}_{8}(dz+{\cal{A}}_{2})
\eea
where ${\cal{A}}_{2}$ is a Kaluza-Klein vector, $z$ is the compactification
coordinate and $x\equiv x^{\mu},\,\mu=0,\dots,7$. The scalar kinetic term
thus reduces as
\be
{\hat e}\;tr\bigg({D}_{{\hat{\mu}}}{\hat{\cal{M}}}
{D}^{{\hat{\mu}}}{\hat{\cal{M}}}^{-1}\bigg)=
e\;tr\bigg({D}_{\mu}{\cal{M}}
{D}^{\mu}{\cal{M}}^{-1}\bigg)+
e\;tr\bigg({D}_{z}{\cal{M}}
{D}^{z}{\cal{M}}^{-1}\bigg).
\ee
The second term in this expression generates the contribution to the
potential
\be
{\cal{V}}^{\prime}_{8}({\cal{M}})={{{m}^{2}}\over4}\,tr\bigg[
({\eta}^{T}{\cal{M}}+{\cal{M}}\eta)
({\eta}^{T}{\cal{M}}^{-1}+{\cal{M}}^{-1}{\eta}^{T})\bigg]{\lambda}_{8}^{2}
e^{2({\alpha}_{8}-{\beta}_{8}){\sigma}_{8}}
\ee
which is easily shown to be
\be
{\cal{V}}^{\prime}_{8}({\cal{M}})={{{m}^{2}}\over2}\,tr\bigg[
{\cal{M}}^{2}-{\bf 1}_{2}\bigg]{\lambda}_{8}^{2}
e^{2({\alpha}_{8}-{\beta}_{8}){\sigma}_{8}}.
\ee
Hence in D=8 the complete scalar potential is
\be
{\cal{V}}_{8}({\cal{M}})={{{m}^{2}}\over8}\,tr\bigg[
{\cal{M}}^{2}-{\bf 1}_{2}\bigg]\bigg[e^{{4\over{\sqrt7}}\varphi}+
{\lambda}_{8}^{2}e^{-2{\beta}_{8}{\sigma}_{8}}\bigg]
e^{2{\alpha}_{8}{\sigma}_{8}}.
\ee
Generalising to a dimensional reduction to $d$ dimensions, it is clear that
the potential will become
\be
{\cal{V}}_{d}({\cal{M}})={{{m}^{2}}\over8}\,tr\bigg[
{\cal{M}}^{2}-{\bf 1}_{2}\bigg]\bigg\{S({\sigma}_{i})\bigg\}
\ee
where
\be
S({\sigma}_{i})=
e^{-2{\beta}_{9}{\sigma}_{9}+2{\sum}_{i=d}^{9}{\alpha}_{i}
{\sigma}_{i}}+
{\lambda}_{8}^{2}
e^{-2{\beta}_{8}{\sigma}_{8}+2{\sum}_{i=d}^{8}{\alpha}_{i}
{\sigma}_{i}}
+{\dots}+
{\lambda}_{d}^{2}
e^{-2{\beta}_{d}{\sigma}_{d}+2{\alpha}_{d}
{\sigma}_{d}}
\ee
and the ${\sigma}_{i}$'s are Kaluza-Klein scalars and the ${\lambda}_{i}$'s
are scalars from the reductions of ${\cal{A}}_{1}$. Clearly
${\cal{V}}_{d}({\cal{M}})\geq 0$ and ${\cal{V}}_{d}({\cal{M}})=0$
when ${\cal{M}}={\bf 1}_{2}$ and all the Kaluza-Klein scalars are arbitrary
constants as in D=9. Thus the $d$-dimensional vacuum is Minkowski spacetime.

As noted above however there may be contributions to the potential from
terms like those of (\ref{termsinB2}). The latter term will yield a
contribution to the D=7 potential of the form
\be
\sim m^{2}e\,e^{{\sum}_{i}{\gamma}_{i}{\sigma}_{i}}({\bf B}_{0}^{T}
{\cal{M}}^{-1}{\bf B}_{0})
\ee
where ${\bf B}_{0}$ is a doublet of scalars from ${\bf B}_{2}$. Clearly
this modification to ${\cal{V}}_{d}({\cal{M}})$ vanishes if ${\bf B}_{0}=0$.
Hence the potential still has a D=7 Minkowski ground state. The first term
in (\ref{termsinB2}) will give a contribution to the potential in D=6.
Because $D{\bf B}_{2}$ takes the following form in D=9
\be
D{\bf B}_{2}=d{\bf B}_{2}+{m\over2}{\eta}{\bf B}_{2}{\cal{A}}_{1}
\ee
the term $\sim m$ in $D{\bf B}_{2}$ will, after reducing to D=6, yield
a contribution to the potential like
\be
\sim m^{2}e\,e^{{\sum}_{i}{\gamma}^{\prime}_{i}{\sigma}_{i}}
({\lambda}_{1}{\bf C}_{0}+{\lambda}_{2}{\bf B}^{\prime}_{0}+
{\lambda}_{3}{\bf B}_{0})^{T}{\cal{M}}^{-1}
({\lambda}_{1}{\bf C}_{0}+{\lambda}_{2}{\bf B}^{\prime}_{0}+
{\lambda}_{3}{\bf B}_{0}).
\ee
Again this will vanish if the various new scalars ${\lambda}_{i},
{\bf C}_{0}, {\bf B}^{\prime}_{0}, {\bf B}_{0}$ coming from the various
reductions of ${\cal{A}}_{1}$ and ${\bf B}_{2}$ vanish.

It is clear that neither the supersymmetry nor the $SO(2)$ gauging
is broken in the Kaluza-Klein reductions thus we obtain a class of $SO(2)$
gauged supergravities with Minkowski ground states in all dimensions
from D=9 to D=4 \footnote{Scherk-Schwarz reductions \cite{SchSch}, have been shown
previously to lead to gauged, massive supergravities with potentials possessing flat directions.}.

After reduction to D=4 the kinetic term for the four-form field strength
can be eliminated in favour of a scalar potential with the form
\be
{\cal{V}}\sim m^{2}e\,e^{\alpha\Phi},
\ee
where $m$ is an integration constant. If $m\neq 0$ then the model will
admit a standard domain wall solution \cite{LPSS}. This reduction plus dualisation is
equivalent to first dualising the four-form in D=5 then performing
a Scherk-Schwarz reduction using a global axionic shift symmetry
\cite{CoLPST}. Clearly one can dimensionally reduce to D=4 and choose not
to dualise the four-form. The resulting model will have a Mink$_4$ vacuum.

\section{Conclusions}

In this paper we have performed a generalised $Sl(2,\bb{R})$ reduction of
type IIB supergravity and obtained the bosonic sector of maximal $SO(2)$
gauged supergravity in nine dimensions. As dimensional reduction preserves
all supersymmetries this D=9 model must admit a supersymmetric extension.

It is well known that by considering the `near horizon geometries' of
supergravity brane (or intersecting brane) solutions in D=10 and 11, one can
infer new reductions leading to gauged supergravities in lower
dimensions and also find new
solutions \cite{boonstra,GT,DGT,cow}. A classic example is the `near horizon geometry' of the D3 brane.
This is $AdS_{5}\times S^{5}$ which implies an $SO(6)$ gauged supergravity in
D=5 with an $AdS_{5}$ ground state. One may ask whether there is a brane
solution of the IIB theory which implies D=9 $SO(2)$ gauged supergravity
with a D=9 Minkowski ground state. The natural place to look is the D7-brane.
One form of this solution was given in \cite{BDRGPT} and takes the following form in
the string frame\footnote{other forms of the solution are given in \cite{GGP,EPZ}}
\bea
ds^2 &=& H^{-{1\over2}}(r) dx_{\mu}dx^{\mu} +
H^{1\over2}(r)\big[ d{z}^2 + dr^2\big] \, \nonumber\\
e^{-\hat\phi} &=& H(r)\nonumber\\
\hat\chi &=& \pm H'(r) z\, .
\eea
where the transverse coordinate $z$ is periodically identified.
${\hat\phi}$ and $\hat\chi$ are the IIB dilaton and axion respectively and
$H(r)$ is an harmonic function of $r$ only. 

It is clear that in the limit of
$r$ tending to zero the metric becomes $Mink_{9}\times S^{1}$, the dilaton
is constant and the field strength of the axion is proportional to the volume form on the $S^{1}$ factor. This is consistent with an $SO(2)$ gauged
supergravity in D=9 with a Minkowski ground state. 

As explained in the introduction, three intersecting M5-branes have a near horizon geometry of $AdS_3{\times}{\bb{E}}^{6}{\times}S^2$ implying an $S^{2}$
reduction of D=11 supergravity is possible. Presumably one obtains the D=9
SO(2) gauged model whose bosonic sector we have presented here. It would
be interesting to confirm this and obtain the fermionic lagrangian
and supersymmetry transformation laws. This would then allow one to give an
M-theory interpretation to the 9D Minkowski and domain wall vacua of section 4.

\section*{Acknowledgements}

The author would like to thank N.D. Lambert, D. Tong and K. Skenderis for many helpful discussions. The author thanks PPARC for financial support and
acknowledges SPG grant PPA/G/S/1998/00613 for further support.


\end{document}